\documentclass[aps,twocolumn,superscriptaddress]{revtex4}
\usepackage{epsfig}
\usepackage{times}
\begin{document}

\title{The evolution of trust and trustworthiness}

\author{Aanjaneya Kumar}
\affiliation{Department of Physics, Indian Institute of Science Education and Research, Dr. Homi Bhabha Road, Pune 411008, India}

\author{Valerio Capraro}
\email{v.capraro@mdx.ac.uk}
\affiliation{Department of Economics, Middlesex University, The Burroughs, London NW4 4BT, U.K.}

\author{Matja{\v z} Perc}
\email{matjaz.perc@gmail.com}
\affiliation{Faculty of Natural Sciences and Mathematics, University of Maribor, Koro{\v s}ka cesta 160, 2000 Maribor, Slovenia}
\affiliation{Department of Medical Research, China Medical University Hospital, China Medical University, Taichung 404, Taiwan}
\affiliation{Complexity Science Hub Vienna, Josefst{\"a}dterstra{\ss}e 39, 1080 Vienna, Austria}

\begin{abstract}
Trust and trustworthiness form the basis for continued social and economic interactions, and they are also fundamental for cooperation, fairness, honesty, and indeed for many other forms of prosocial and moral behavior. However, trust entails risks, and building a trustworthy reputation requires effort. So how did trust and trustworthiness evolve, and under which conditions do they thrive? To find answers, we operationalize trust and trustworthiness using the trust game with the trustor's investment and the trustee's return of the investment as the two key parameters. We study this game on different networks, including the complete network, random and scale-free networks, and in the well-mixed limit. We show that in all but one case the network structure has little effect on the evolution of trust and trustworthiness. Specifically, for well-mixed populations, lattices, random and scale-free networks, we find that trust never evolves, while trustworthiness evolves with some probability depending on the game parameters and the updating dynamics. Only for the scale-free network with degree non-normalized dynamics, we find parameter values for which trust evolves but trustworthiness does not, as well as values for which both trust and trustworthiness evolve. We conclude with a discussion about mechanisms that could lead to the evolution of trust and outline directions for future work.
\end{abstract}

\maketitle

\section{Introduction}
While we live in a time where the average individual is much healthier and safer than ever before \cite{pinker_11, pinker2018enlightenment}, we are also daunted by several political conflicts, health threats, and extreme poverty in many parts of the world. New innovations and technological breakthroughs often seem to promise a better tomorrow, but the privileges remain restricted to only a tiny fraction of the population. While the issues of equality and egalitarianism are certainly multi-faceted, it is clear that solutions would require us to act prosocially, giving up parts of our personal benefits to help others. But the caveat is that behaving prosocially is costly and not optimal for the individual, and thus will not present itself unless additional mechanisms are at play. No wonder, thus, that understanding the mechanisms that favor the evolution of prosocial behavior has been declared one of the greatest challenges of the 21st century, and that scholars from disciplines as diverse as sociology, psychology, anthropology, economics, biology, and physics have tried to solve the puzzle \cite{hardin1968tragedy, trivers1971evolution, axelrod1981evolution, milinski2002reputation, boyd2003evolution, nowak2006five, hilbe2010incentives, capraro2013model, rand2013human, perc_pr17}.

An exciting development during the past two decades has been the coming of age of network science \cite{barrat_08, newman_10, estrada2012structure, barabasi_16}, which combined with other methods of statistical physics \cite{stanley_71, liggett_85, landau_00}, has reached a level of maturity that allows us to tackle some of the greatest challenges of our time. The study of social dynamics \cite{castellano_rmp09}, traffic \cite{helbing2001traffic}, crime \cite{orsogna_plr15}, epidemic processes \cite{pastor_rmp15}, climate inaction \cite{pacheco_plrev14}, and vaccination \cite{wang_z_pr16} are all examples of this exciting development, which can be put under the umbrella of social physics \cite{perc_srep19}. Prosocial behavior is no exception either and, in particular, the Monte Carlo method for the simulation of evolutionary games and related models on networks has been used prolifically to shed light on the mechanism that may promote it. Most previous works have focussed on three kinds of prosocial behavior, namely cooperation \cite{santos2005scale, pacheco2006coevolution, gomez2007dynamical, ohtsuki2007breaking, lee2011emergent, tanimoto_pre12, wang_plr15, javarone_epjb16, amaral2018heterogeneous2, vilone2018hierarchical}, strategic fairness \cite{szolnoki2012defense, page2000spatial, kuperman2008effect, eguiluz2009critical, da2009statistical, deng2011coevolutionary, gao2011coevolutionary, szolnoki2012accuracy}, and altruistic punishment \cite{helbing_ploscb10, szolnoki_prx13, szolnoki_prx17}.

However, recent empirical research in experimental economics and psychology suggests that two of these behaviors -- cooperation and altruistic punishment -- can be seen as a special form of a more general class of behavior, namely moral behavior \cite{dal2014right, biziou2015does, kimbrough2016norms, eriksson2017costly, capraro2018right, capraro2019increasing}. This observation opens up the possibility of using the same methods that have been used to study the evolution of cooperation and altruistic punishment to effectively study also the evolution of other types of moral behavior \cite{capraro_fp18}. Following this idea, recent work has explored the evolution of lying and found a number of intriguing conditions for the evolution of truth-telling \cite{capraro2019evolution, capraro2020lying}. Therefore, motivated by the success of this new line of work, we here apply the same methods to study the evolution of trust and trustworthiness.

While the precise definition of trust and trustworthiness depends upon the specific context in which it is being used, a general feature that it exhibits is the willingness of an agent -- the trustor -- to act in such a way that she is placed in a vulnerable situation with respect to another agent -- the trustee, especially when the trustor has no direct ability to monitor the trustee's actions. Thus, trust invariably involves putting oneself in a vulnerable situation in the hope of high returns. High returns that, in the absence of any mechanism to enforce the reciprocation of the trust, might never come, because the trustee can maximize his gain by simply walking away with the profit obtained by betraying the trustor. Knowing this, the trustor should not trust in the first place. Therefore, both trust and trustworthiness go against the assumptions of narrow self-interest. They in fact correlate with several measures of morality, including cooperation and altruism \cite{peysakhovich2014humans}. Moreover, trustworthiness, in the form of `returning favors', has been recently found to be a universal moral rule across 60 societies around the world \cite{curry2019good}. Yet, despite the fact that both trust and trustworthiness go against the assumption of narrow self-interest, we see them in action everyday -- from travelers preferring to look for accommodation through Airbnb rather than spending money on booking a hotel room, to computers in a network deciding to receive information from a source outside of their network. Trust and trustworthiness are ubiquitous in our society, which suggests that, in reality, some mechanisms that favor the evolution of trust and trustworthiness must be at play. The question is which are these mechanisms?

Real interactions do not happen in a vacuum, nor are they random. They are inherently limited to a subset of the population. Some interactions are more frequent than others, and some individuals have many more contacts than others. We are far more likely to interact with friends, family members, and co-workers, than we are with random people. The very fact that interactions are structured has been shown to promote cooperation, along the logic of network reciprocity: cooperators can form clusters to protect themselves from the invasion of defectors \cite{nowak2006five}. Similarly, it has been shown that spatial structure favors the evolution of fairness and altruistic punishment \cite{page2000spatial, helbing_ploscb10}, as well as the evolution of truth-telling, at least in some cases \cite{capraro2020lying}. In this paper, we take inspiration from this line of research and we ask whether network reciprocity promotes also the evolution of trust and trustworthiness.

The plan of the paper is as follows: in Section~\ref{trust game}, we will describe the trust game and give a brief overview of the research attention that it has attracted since its introduction. In Section~\ref{monte carlo}, we will provide a description of the Monte Carlo method used to numerically evaluate the stationary state frequencies of different strategies in the trust game, played in well-mixed as well as in networked populations. We will present our results in Section~\ref{results}, and we will end with a summary and outlook for future research in Section~\ref{conclusion}.

\section{Methods}
\label{methods}

\subsection{The trust game}
\label{trust game}

Berg, Dickhaut and McCabe \cite{berg1995trust} proposed the trust game in 1995 as an elegant way to measure trust and trustworthiness between two agents. Player $A$ (the trustor) is initially given some amount of money, normalized to $1$. In the first step of the game, player $A$ can choose to \emph{trust} player $B$ (the trustee) and transfer a proportion $x\in [0,1]$ of her endowment to player $B$. $A$ transfer of $x=0$ corresponds to player $A$ choosing to not trust $B$ and to walk away with her money; in this case, the game ends. Instead, if $A$ transfers some amount $x>0$ to $B$, the amount of money transferred to player $B$ is tripled (i.e. $B$ gets $3x$ units of money while $A$ is left with $1-x$) and the game continues. In the second step, player $B$ chooses a fraction $r \in [0,1]$ of the money he possesses to return to player $A$. This marks the end of the game. Therefore, the final payoffs of player $A$ and player $B$ are, respectively, $1-x+3xr$ and $3x(1-r)$ units of money.

In a one-shot anonymous trust game, it is clear that a self-interested player $B$ has no incentive to return any amount of money to player $A$. This backward induction argument suggests that the best strategy for player $A$ would be to not trust player $B$. However, experimental research has repeatedly reported that a significant proportion of people choose to transfer a non-zero amount of money to their co-player and a substantial amount of money is also returned \cite{berg1995trust, glaeser2000measuring, bohnet2004trust, malhotra2004trust, kosfeld2005oxytocin, fehr2009economics, johnson2011trust}. Importantly, this behaviour cannot be explained by lack of comprehension \cite{ortmann2000trust} or risk aversion \cite{fetchenhauer2009people, dunning2012trust, corgnet2016trust, ben2010trusting}. Specifically, one observes trust and trustworthiness also among experimental participants who have a clear understanding of what their payoff-maximising strategy is. And trust does not seem to be driven by risk seeking: many individuals who choose to trust in the trust game are averse to taking the risk in an equivalent lottery. In summary, the empirical literature on the trust game provides a clear indication that, while trust and trustworthiness go against monetary payoff maximisation, they often emerge. In order to better comprehend the origin and evolution of trust and trustworthiness, experimental studies need to be complemented with extensive numerical simulations which can help us shed light on when and how trust can be selected in a population and what role does the structure of the population has on its evolution.

At this stage, one might wonder whether trust and trustworthiness are fundamentally different from other forms of social behaviour that have been studied with methods of statistical physics. The answer is positive. This is easy to see in the case of the ultimatum game (used to measure strategic fairness and altruistic punishment) and the sender-receiver game or other deception games (used to measure lying), because they, compared to the trust game, have a completely different payoff structure and set of Nash equilibria. If anything, the trust game \emph{looks} similar to the prisoner's dilemma, the symmetric game in which both players have to decide whether to cooperate or defect: cooperation means paying a cost to give a greater benefit to the other player; defecting means doing nothing. Although the trust game and the prisoner's dilemma might superficially look similar, they are actually fundamentally different. Not only the trust game differs from the prisoner's dilemma on the technical fact that the latter is symmetric, while the former is not, but, more crucially, it fundamentally differs in the evolutionary patterns that it generates, as we will now show.

\subsection{The Monte Carlo method}
\label{monte carlo}

In the trust game, the amount $x$ that player $A$ transfers to player $B$ is considered as an individual measure of trust, whereas the fraction $r$ that player $B$ returns to player $A$ is taken as a measure of trustworthiness. While in theory any amount of trust $x\in[0,1]$ and any amount of trustworthiness $r\in[0,1]$ can be possible, in practice, people in the position of player $A$ often have a binary decision to make, whether to trust or not to trust people in the position of Player $B$; similarly, people in the position of player $B$ often have a binary decision to make, whether to return a previously agreed amount of money or not \cite{ermisch2009measuring, espin2016heterogeneous}. We follow this line of work and we also consider a binary version of the trust game, in which player $A$ can either choose to trust (T) or not trust (N), whereas player $B$ can either reciprocate (R) player $A$'s trust or betray (B). This yields a payoff bimatrix:
\begin{center}
\begin{tabular}{ |c|c|c| }
 \hline
  & T & N \\
 \hline
 R & $1+(3r-1)x, 3(1-r)x$ & $1,0$ \\
 B & $1-x, 3x$ & $1,0$  \\
 \hline
\end{tabular}
\end{center}
A particularly interesting case is when $x=1$ and $r=0.5$, corresponding to the case in which the trustor invests all her money which is normalized to $1$ and the amount that the trustee can return corresponds to an equal split between them.

We carried out simulations of the trust game for well mixed populations as well as several network structures (hexagonal, square, and triangular lattices, as well as random networks and scale-free networks) using the Monte Carlo (MC) method. For a well-mixed population with $N$ players, the following are the elementary steps: The simulation starts by randomly distributing the four strategies (T,R), (T,B), (N,R) and (N,B) among $N$ agents. Two players $P_1$ and $P_2$ are then randomly picked and they play the trust game with four randomly chosen neighbors. We note that since these players are picked randomly without restricting the selection to nearest neighbors or linked players in a network, the procedure thus yields well-mixed conditions. In each of the eight games, the roles of players are assigned randomly. $P_1$ and $P_2$ collect payoffs $\Pi_{P_1}$ and $\Pi_{P_2}$, respectively. Then Player $P_2$ copies the strategy of player $P_1$ with probability
\begin{equation}
    w=1/(1+\exp(\Pi_{P_2}-\Pi_{P_1})/K)
\end{equation}
where we choose $K=0.1$. This step is repeated $N$ times, which by definition completes one full MC step \cite{binder_88}. During the repetition of many full MC steps, every player will thus (since $N$ is also the population size) have a chance once, on average, to change its strategy for each full Monte Carlo step that is made. Indeed, in our simulations, we have performed the MC method for up to $10000$ full MC steps. This completes one realization. We conducted $5000$ realizations, using randomized initial conditions and the evolution described in the Results section is obtained by averaging over these realizations.

For structured populations, we introduce the constraint in the above mentioned elementary steps that $P_1$ and $P_2$ must necessarily be neighbours, or directly linked players in a network. In the case of heterogeneous networks, we use two different imitation rules -- the normalized and the unnormalized replicator dynamics. In the unnormalized dynamics,
the probability with which $P_2$ replicates the strategy of $P_1$ is as before:
\begin{equation}
    w=1/(1+\exp(\Pi_{P_2}-\Pi_{P_1})/K).
\end{equation}
The normalized replication probability differs from the unnormalized one in that payoffs are scaled down by the degree of the players, that is,
\begin{equation}
w=1/(1+\exp(\Pi'_{P_2}-\Pi'_{P_1})/K),
\end{equation}
where $\Pi'_{P_i}=\frac{\Pi_{P_i}}{k_i}$ and $k_i$ is the degree of player $i$.
In practice, this means that, in the normalized replicator dynamics, players take into account their degree and the degree of other players; whereas, this does not happen in the unnormalized dynamics. Both  dynamics are useful in their domain of applicability. The normalized replicator dynamics is useful in situations in which the degrees are visible, and the individuals make a fair comparison with their neighbours -- this often happens online, as social media allows people to visualize the connections of an agent with other agents. The unnormalized dynamics is useful in situations in which the imitation of strategies happens on the basis of how well the other player is doing, without taking into account the number of connections they have -- for example, people trying to mimic the habits of successful people without accounting for the number of resources that they have at their disposal compared to the person they are copying.

\section{Results}
\label{results}

\subsection{Well-mixed populations}

\begin{figure}
\centerline{\epsfig{file=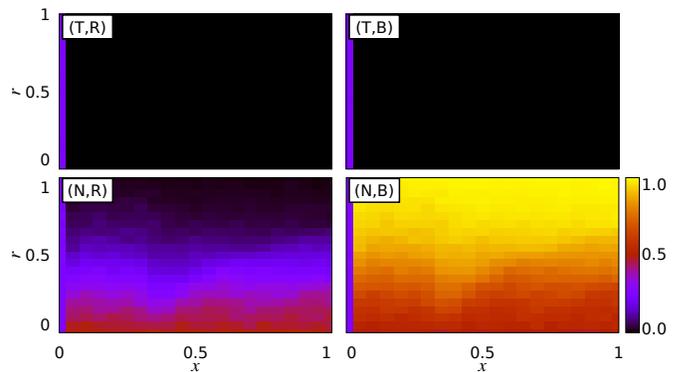,width=9cm}}
\caption{The stationary density of the four strategies plotted on a $20\times20$ grid of $(x,r)$ values with both $x$ and $r$ ranging from 0 to 1. Well-mixed population.}
\label{fig:mesh}
\end{figure}

We first report the final densities of the four strategies (T,R), (T,B), (N,R) and (N,B), as a function of the parameters $x$ and $r$, in a well-mixed population consisting of $500$ agents. Figure \ref{fig:mesh} highlights that, in this case, trust does not evolve, as both the strategies (T,R) and (T,B) appear with density 0 at the steady state, irrespective of the parameters $x$ and $r$. By contrast, the final density of trustworthiness highly depends on the parameter $r$, while being insensitive to the parameter $x$. Specifically, for each $x$, the prevalence of trustworthiness is about 50\% for very small values of $r$, and then monotonically decreases as $r$ increases.

In order to gain a better understanding of the evolution of trust and trustworthiness, we also conducted several simulations to study the time evolution of the frequencies. We do not report the outputs in the figures, as they are all very similar and certainly not surprising. For example, for $x=1$ and $r=0.5$, consistent with Figure \ref{fig:mesh}, we found that the frequencies of the strategies (T,R) and (T,B) go to zero after about 40 MC steps. On the other hand, the strategy (N,B) survives with very high frequency, around 85\%. The remaining frequency, about 15\%, is taken by the remaining strategy, (N,R).

\subsection{Lattices}

\begin{figure}
\centerline{\epsfig{file=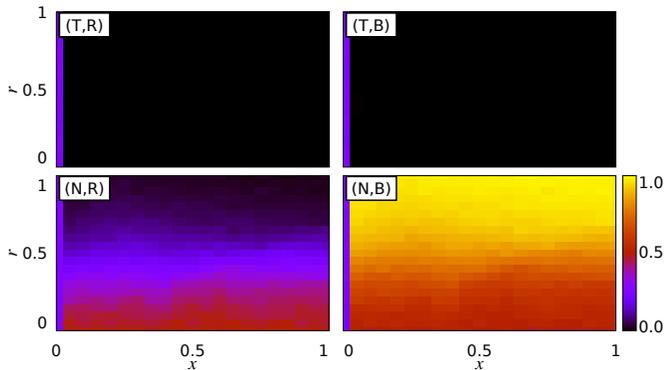,width=9cm}}
\caption{The stationary density of the four strategies plotted on a $20\times20$ grid of $(x,r)$ values with both $x$ and $r$ ranging from 0 to 1. Square lattice.}
\label{fig:rs_square_lattice}
\end{figure}

\begin{figure}[b]
\centerline{\epsfig{file=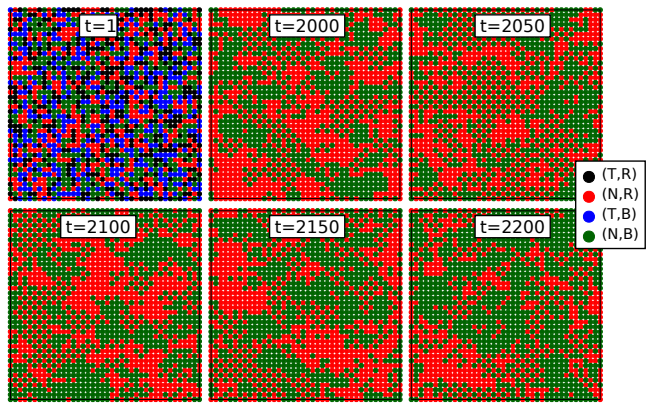,width=8.5cm}}
\caption{Snapshots of the evolution of strategies in the trust game played on a $40\times40$ square lattice. In most realizations of the game, the entire system enters an absorbing state. However, in a few realizations, (N,B) and (N,R) both survive for long time. We have picked one such realization for representation purposes.}
\label{fig:snap}
\end{figure}

To understand the role of the spatial structure on the evolution of trust and trustworthiness, we simulate the trust game on different lattices.  Figure \ref{fig:rs_square_lattice} reports the stationary densities of the four strategies as a function of $x$ and $r$ on the square lattice. It is immediately evident that the trends in the evolution of different strategies remain the same, compared to the well-mixed populations (Figure \ref{fig:mesh}). We obtain similar trends in the case of the triangular and the hexagonal lattices (figures reported in the supplementary information) with the results differing only by a small numerical value.

In order to provide further evidence that the lattice structure has very little effect on the evolution of trust and trustworthiness, we also conducted several simulations to study the time evolution of the four strategies in the three lattices (figures not reported in the paper). Consistent with the results mentioned above, we found that the time evolution in the lattice is very similar to the well-mixed case. For example, for $x=1$ and $r=0.5$, we found that the frequencies of (T,R) and (T,B) quickly go to zero, whereas the final density of (N,R) is slightly larger that it was in well-mixed populations, but the numerical difference is very small (around 5\%); consequently, the final density of (N,B) is slightly smaller in the lattices than it was in well-mixed populations.

To better understand the spatial evolution of the strategies, Figure \ref{fig:snap} presents snapshots of the game on a square lattice at late times. In most realizations of the game, the whole population adopts a single strategy and the system enters an absorbing state. However, in a few realizations, (N,B) and (N,R) both survive for long time. We have picked one such realization for representation purposes. It is clearly seen that the two surviving strategies tend to cluster together, forming metastable clusters. It is also noticed that, apart from clusters, there are patches where the two surviving strategies appear alternatively.

\subsection{Random and scale-free networks}

\begin{figure}
\centerline{\epsfig{file=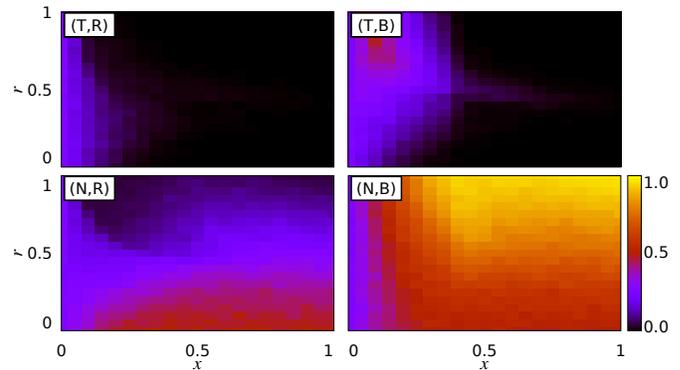,width=9cm}}
\caption{The stationary density of the four strategies plotted on a $20\times20$ grid of $(x,r)$ values with both $x$ and $r$ ranging from 0 to 1. Scale free network (unnormalized replicator dynamics).}
\label{fig:scale_free_unnormalized}
\end{figure}

To investigate the role of the spatial structure on the evolution of trust and trustworthiness further, we simulate the trust game on a scale-free network generated by the Barab{\'a}si-Albert algorithm and an Erd{\H{o}}s-R{\'e}nyi random network, with $500$ agents each (both with an average degree close to 10). Since these networks are not regular, we consider both the normalized and the unnormalized replicator dynamics.

In the case of random networks, we obtain results very similar to the well-mixed populations and the three lattices. This holds using both the normalized and unnormalized dynamics (see supplementary information) and it provides further evidence that spatial correlations alone do not lead to the evolution of trust.

\begin{figure}
\centerline{\epsfig{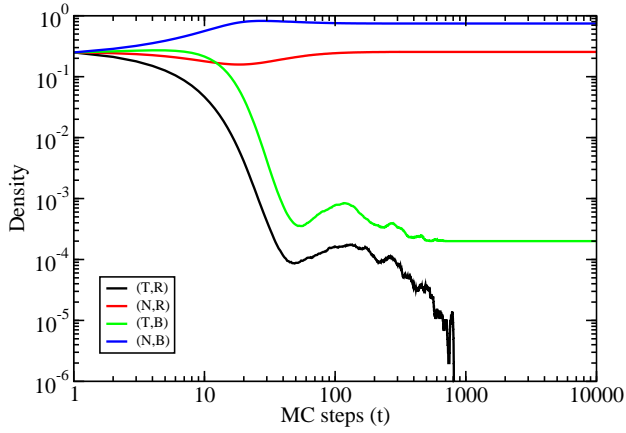}}
\vspace{0.8cm}
\centerline{\epsfig{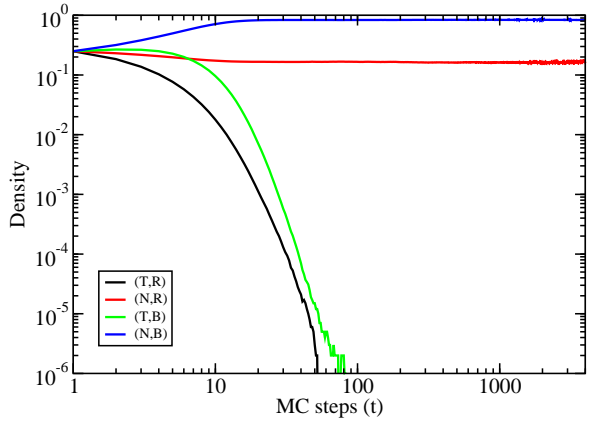}}
\caption{Time evolution of the four strategies in a scale free network for values $x=1$ and $r=0.5$, for the unnormalized (top) and normalized (bottom) replicator dynamics.}
\label{fig:scale_free_x1_r05}
\end{figure}

To study the effects of heterogeneity in the network of contacts, we study the trust game on scale-free networks. In this case, evolution turns out to be more nuanced and interesting, as it appears to depend on the choice of replicator dynamics. When agents imitate other agents using the normalized replicator dynamics, trust does not evolve and the results are virtually the same as in the previous cases. However, when agents imitate other agents using the unnormalized replicator dynamics, we observe rich behaviour in terms of the evolution of different strategies.  Figure \ref{fig:scale_free_unnormalized} is a heatmap of the steady state density of the four strategies, and it shows clear differences with Figure \ref{fig:mesh}. The strategy (T,R), which in the other cases never evolved, this time evolves for small values of $x$. Similarly, the strategy (T,B), which generally vanished in the other cases, now evolves for $x<0.5$. The difference is particularly evident for $x$ small and $r$ large, where there appears to be an island in which (T,B) actually evolves with frequency close to 50\%.

\begin{figure}
\centerline{\epsfig{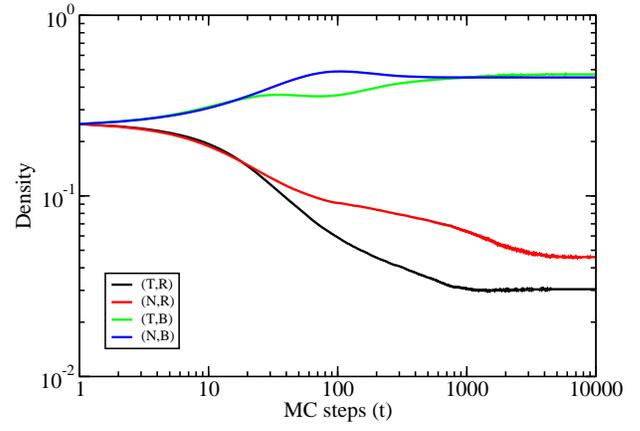}}
\vspace{0.8cm}
\centerline{\epsfig{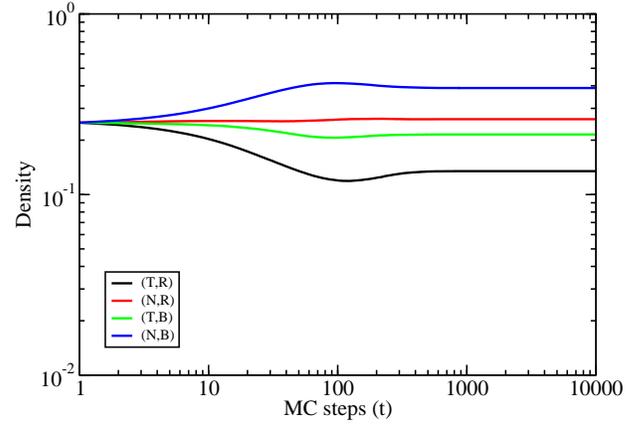}}
\caption{Time evolution of the four strategies in a scale-free network (unnormalized replicator dynamics) for values $x=0.1$ and $r=0.8$ (top) and for $x=0.1$ and $r=0.3$ (bottom).}
\label{fig:scale_free_differences}
\end{figure}

To have a better understanding of these differences, we next explore the time evolution of the four strategies in three prototypical cases, one in which we expect virtually no differences compared to the well-mixed case ($x=1$ and $r=0.5$) and two in which we expect large differences ($x=0.1, r=0.8$ and $x=0.1,r=0.3$). Note indeed that Figure \ref{fig:scale_free_unnormalized} suggests that, for $x=1$ and $r=0.5$, the final densities according to the unnormalized replicator dynamics should be very similar to those according to the normalized replicator dynamics, which are in turn very similar to those in well-mixed populations. By contrast, we chose the values $x=0.1$ and $r=0.8$ to illustrate the evolution in correspondence to the island described above where we expect trust but not trustworthiness to evolve. And we chose the values $x=0.1$ and $r=0.3$ to illustrate a situation in which we expect both trust and trustworthiness to evolve.

Figure \ref{fig:scale_free_x1_r05} reports the time evolution of the four strategies for $x=1$ and $r=0.5$ using the unnormalized dynamics (top panel) and the normalized dynamics (bottom panel). As expected, the bottom panel is virtually identical to the well-mixed population (not reported in the figures, but discussed earlier in the text). The top panel differs from the bottom panel only in a very small detail: the strategy (T,B) evolves with a very small frequency.

\begin{figure}
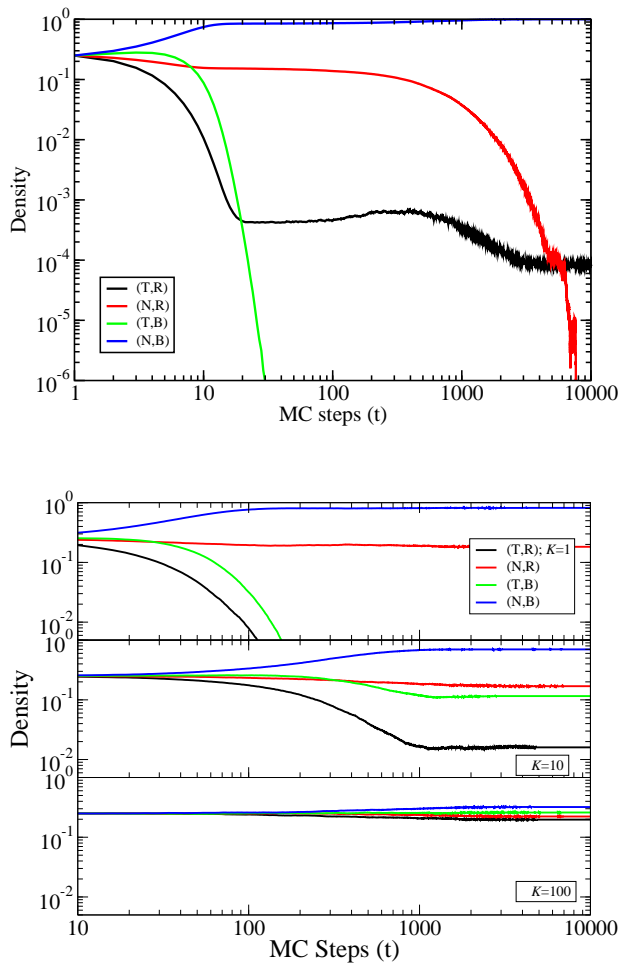

\centerline{\epsfig{file=figure7a.eps,width=8.1cm}}
\vspace{0.8cm}
\centerline{\epsfig{file=figure7b.eps,width=8.1cm}}
\caption{(Top panel) The evolution of the four strategies as a function of time in the case of a well mixed population after adding a `good samaritan' agent that always chooses the strategy (T,R). We look at the evolution of the $499$ agents in the population (excluding the good samaritan) for the values $x=1$ and $r=0.5$. (Bottom panel) Time evolution of the four strategies (without any good samaritan) for $x=1$, $r=0.5$, and varying levels of noise ($K=1, 10, 100$).}
\label{good}
\end{figure}

Figure \ref{fig:scale_free_differences} (top panel) reports the evolution of the four strategies for $x=0.1$ and $r=0.8$, only for the unnormalized replicator dynamics. As expected, this time we see very large differences compared to the normalized replicator dynamics, which we do not report in the paper, being virtually identical to Figure \ref{fig:scale_free_x1_r05}. In particular, the biggest difference can be observed in the evolution of the strategy (T,B). In the previous case ($x=1$, $r=0.5$), this strategy almost vanished when agents imitate other agents using the unnormalized replicator dynamics, and completely vanished when they used the normalized replicator dynamics. In stark contrast, the strategy (T,B) now evolves with frequency close to 50\%. Another difference can be noticed in the case of the strategy (T,R). This strategy vanished in all the earlier cases. By contrast, it now survives, although with a very small probability around 2\%. Finally, Figure \ref{fig:scale_free_differences} (bottom panel) reports the evolution of the four strategies for $x=0.1$ and $r=0.3$, again only for the unnormalized replicator dynamics. As expected, this time the strategy (T,R) evolves with a non-negligible frequency around 15\%. The strategy (N,R) evolves with an even higher frequency around 25\%, compared to the 5\% for $x=0.1$ and $r=0.8$. These increases in frequency come mainly at the expenses of the strategy (T,B), which, for $x=0.1$ and $r=0.8$ evolved with very high frequency (about 45\%), whereas it now evolves only with frequency around 20\%; and to a lesser extent at the expenses of the strategy (N,B), which, for $x=0.1$ and $r=0.8$ evolved with frequency around 45\%, whereas it now evolves with frequency below 40\%.

\section{Discussion}
\label{conclusion}

We have used the Monte Carlo method to study the evolution of trust and trustworthiness in well-mixed populations, three different types of lattices, random networks and scale-free networks. Since the latter two networks are not regular, in these cases we have studied the evolution of trust and trustworthiness both when agents imitate other agents by taking into account their degree (normalized replicator dynamics) and when they do not (unnormalized replicator dynamics). As a measure of trust and trustworthiness, we have used a binary version of the trust game \cite{berg1995trust}. The choice made by player 1 (the trustor) was taken as a measure of trust; the choice made by player 2 (the trustee) was taken as a measure of trustworthiness. We parameterized the game through two parameters: $x\in[0,1]$ describes the amount of money that the trustor can send to the trustee; $r\in[0,1]$ represents the proportion of the amount received by the trustee that the he can return to the trustor.

Our exploration provided evidence of several results. First, in well-mixed populations, trust never evolves, whereas the evolution of trustworthiness depends monotonically decreasingly on $r$ (and shows very little dependence on $x$). Second, to understand the effects of spatial correlations on the evolution of different strategies, we simulated the trust game on homogenous and heterogenous networks. On lattices, random networks (using both the imitation dynamics), and scale-free networks with normalized replicator dynamics, we observe that that trust does not evolve, and in these cases, the final densities of the four strategies are very similar to the corresponding final densities in well-mixed populations. This conclusively points to the fact that solely spatial structure does not lead to the evolution of trust. Third, scale-free networks with unnormalized replicator dynamics give rise to the most nuanced evolution: for small values of $r$ and $x$, both trust and trustworthiness evolve, although with a relatively small frequency around 15\%; for small values of $x$ and large values of $r$, trust evolves with a relatively large frequency around 50\%, but this time trustworthiness does not evolve. These results can readily be compared to the evolutionary prisoner's dilemma on scale-free networks \cite{masuda_srep12} with normalized and unnormalized replicator dynamics where the evolution of cooperation is possible in both cases. The heterogeneous scale-free network provides a mechanism for the survival of cooperators up to larger values of temptation to defect, when compared to well-mixed populations. However, it is interesting to note that when the payoffs of an individual are normalized by their degrees, the fraction of surviving cooperators is significantly lesser. Our results hint that while heterogeneity also provides a route for the evolution of trust, it only does so when the payoff of the players is accumulated over all its interactions with its neighbours, and not averaged over them, like in the normalized dynamics.

In sum, we have operationalized trust and trustworthiness using the trust game with the trustor's investment and the trustee's return of the investment as the two key parameters and we have studied their evolution in a number of networks and our results have shown that trust and trustworthiness very rarely evolve in these networks, and even more rarely do they do it together: when trustworthiness evolves, then trust does not; when trust evolves, trustworthiness does not. Only in a relatively small region (both $r$ and $x$ small) and only in the case of scale-free networks and unnormalized replicator dynamics, the strategy (T,R) evolved with a non-negligible, although still relatively small (around 15\%) probability.

This is the first systematic study on the evolution of trust and trustworthiness on networks. Most previous work applied the Monte Carlo method to study the evolution of cooperation in the prisoner's dilemma \cite{santos2005scale, pacheco2006coevolution, gomez2007dynamical, ohtsuki2007breaking, lee2011emergent, tanimoto_pre12, wang_plr15, javarone_epjb16, amaral2018heterogeneous2, vilone2018hierarchical}, the evolution of strategic fairness and altruistic punishment in the ultimatum game \cite{szolnoki2012defense, page2000spatial, kuperman2008effect, eguiluz2009critical, da2009statistical, deng2011coevolutionary, gao2011coevolutionary, szolnoki2012accuracy}, and the evolution of truth-telling in the sender-receiver game \cite{capraro2019evolution, capraro2020lying} or in other deception games \cite{grafen1990biological, catteeuw2014evolution, han2013good, han2015avoiding, pereira2017evolution}. These games are fundamentally different from the trust game used in the current analysis. The trust game is obviously different from the sender-receiver game and the ultimatum game, because they have completely different strategic structure and, consequently, sets of equilibria. But it is also different from the prisoner's dilemma: while this game is symmetric, the trust game is not. This is probably the reason that leads to the fact that, in general, the spatial structure favors the evolution of cooperation in the prisoner's dilemma, while having very little effect on the evolution of trust and trustworthiness. A handful of papers have studied the evolution of trust and trustworthiness using the trust game or some variants thereof. However, most of these works focused on well-mixed populations \cite{mcnamara2009evolution, manapat2013information, manapat2012delayed, abbass2015n, rauwolf2018expectations}. These works typically show that, with no additional mechanisms, such as choice visibility, trust and trustworthiness do not evolve in well-mixed populations. Our findings are thus in line with this preceding research. A variant of the trust game has also been studied on networks, however the analysis was mainly focused more on group effects, and on one specific network -- the email network of a university in Tarragona \cite{chica2019effects}.

The fact that the spatial structure, apart from one special case, does not promote the evolution of trust and trustworthiness together with the observation that, in reality, we do see a lot of trust and trustworthiness, generates the following question: What mechanisms promote the evolution of trust and trustworthiness? In Figure 7, the top panel is a plot of the evolution of the strategies of $499$ agents for values $x=1$ and $r=0.5$ in a well-mixed population of $500$ agents where the excluded individual is a `good samaritan' who always chooses the strategy (T,R). We emphasize that the plot only considers the evolution of the strategies of the $499$ agents which does not include the good samaritan. It can be seen that a finite, albeit small fraction of trustors, survive in the stationary state upon the inclusion of a single good samaritan agent at a value of $x$ and $r$ where previously trust did not evolve. It is well known that zealots can drive the evolution of cooperation in the prisoner's dilemma game \cite{santos_jeb06} and a detailed study of the effects of good samaritans on the evolution of trust and other moral behaviours provides an interesting avenue for future research. The bottom panel of Figure 7 explores the influence of noisy imitation on the dynamics. Noise can be interpreted as the lack of perfect information about the payoffs of other people, or as sub-optimal decision making. We show a comparison of the evolution of increased noise in the imitation process for $x=1$, $r=0.5$, and $K=1, 10$ and $100$). It is expected in the limit of $K \rightarrow \infty$ that each strategy survives with equal probability as the dynamics is random. However, even at $K=10$, we can see that trust evolves to steady state density of around $10$\% and the strategy (T,R) which accounts for trusting, and trustworthy individuals also evolve to a final density of around $1$\%. Studying further the effects of noisy imitation as a function of the parameters of the game could lead us to novel insights.

Several other mechanisms could be responsible for the evolution of trust \cite{andras2018trusting}. Possible candidates could be reward and punishment as well as apology, forgiveness, and emotions such as guilt. We know that these mechanisms promote the evolution of cooperation \cite{rand_s09, gurerk2006competitive, szolnoki_njp12, milinski2006stabilizing, martinez2015apology, capraro2016partner, melo2016people, martinez2017agreement, pereira2017social, fang2019synergistic}. Along similar lines, it is possible that they also promote the evolution of trust and trustworthiness. In fact, in reality, we know that, for example, online transactions, which are fundamentally based on a relationship of trust, are supported by rating systems that provide a measure of the trustworthiness of the agents. Therefore, it is likely that the presence of a reputation mechanism promotes the evolution of trust. Following recent works of Fudenberg and Imhof \cite{fudenberg_jet06} and Veller and Hayward \cite{veller2016finite}, it would also be interesting to study the problem where not only do the agents evolve using imitation, but also can spontaneously mutate and adopt different strategies. Additionally, we notice that our results were obtained on particular networks and imitation rules; it is possible that other networks and/or other imitation rules lead to the evolution of trust and trustworthiness. Finally, individual differences for example in gender, age, dominance status, number of neighbours, kinship, which are well-known to affect cooperative and altruistic behaviour \cite{taylor2000overlapping,bao2012reproductive,rand2016social,rand2017social,branas2018gender,vilone2018hierarchical}, can also affect the evolution of trust and trustworthiness. Future work should explore these possibilities.

\section*{Acknowledgments}
Aanjaneya Kumar would like to acknowledge the Prime Minister's Research Fellowship for financial support, and thank Shraddha Pathak and Sandeep Chowdhary for many useful discussions. Matja{\v z} Perc was supported by the Slovenian Research Agency (Grant Nos. J4-9302, J1-9112, and P1-0403).

\providecommand{\noopsort}[1]{}\providecommand{\singleletter}[1]{#1}%

\end{document}